\documentclass[twocolumn,showpacs,amsmath,amssymb,superscriptaddress]{revtex4}

\usepackage{graphicx}
\usepackage{dcolumn}
\usepackage{bm}
\usepackage{amssymb}

\newcommand{\pT}{\mbox{$p_\perp$}}
\newcommand{\ET}{\mbox{$E_\perp$}}

\newcommand{\ie}{\textrm{i.e.}}

\newcommand{\vs}{\textrm{vs.}}
\newcommand{\sqrts}{\mbox{$\sqrt{s}$}}

\newcommand{\jpsi}{\mbox{$J/\psi$}}
\newcommand{\upsi}{\mbox{$\Upsilon$}}
\newcommand{\upsione}{\mbox{$\Upsilon(\textrm{1S})$}}
\newcommand{\upsitwo}{\mbox{$\Upsilon(\textrm{2S})$}}
\newcommand{\upsithree}{\mbox{$\Upsilon(\textrm{3S})$}}
\newcommand{\gev}{\mbox{$\mathrm{GeV}$}}

\newcommand{\mevcc}{\mbox{$\mathrm{MeV}/c^2$}}
\newcommand{\gevc}{\mbox{$\mathrm{GeV}/c$}}
\newcommand{\mevc}{\mbox{$\mathrm{MeV}/c$}}
\newcommand{\gevcc}{\mbox{$\mathrm{GeV}/c^2$}}

\newcommand{\pp}{\mbox{$p+p$}}
\newcommand{\AuAu}{\mbox{Au+Au}}
\newcommand{\dAu}{\mbox{$d$+Au}}

\newcommand{\Tc}{\mbox{$T_c$}}
\newcommand{\epluseminus}{\mbox{$e^+e^-$}}
\newcommand{\invpb}{\mbox{$\mathrm{pb}^{-1}$}}
\newcommand{\bbbar}{\mbox{$b$-$\bar{b}$}}

\begin{document}

\preprint{Version 5.1}

\title{\protect\bm{\upsi} cross section in \protect\bm{$p+p$} collisions at \protect\bm{$\sqrt{s} = 200$} GeV} 

\affiliation{Argonne National Laboratory, Argonne, Illinois 60439,
USA} \affiliation{University of Birmingham, Birmingham, United
Kingdom} \affiliation{Brookhaven National Laboratory, Upton, New
York 11973, USA} \affiliation{University of California, Berkeley,
California 94720, USA} \affiliation{University of California, Davis,
California 95616, USA} \affiliation{University of California, Los
Angeles, California 90095, USA} \affiliation{Universidade Estadual
de Campinas, Sao Paulo, Brazil} \affiliation{University of Illinois
at Chicago, Chicago, Illinois 60607, USA} \affiliation{Creighton
University, Omaha, Nebraska 68178, USA} \affiliation{Czech Technical
University in Prague, FNSPE, Prague, 115 19, Czech Republic}
\affiliation{Nuclear Physics Institute AS CR, 250 68
\v{R}e\v{z}/Prague, Czech Republic} \affiliation{University of
Frankfurt, Frankfurt, Germany} \affiliation{Institute of Physics,
Bhubaneswar 751005, India} \affiliation{Indian Institute of
Technology, Mumbai, India} \affiliation{Indiana University,
Bloomington, Indiana 47408, USA} \affiliation{University of Jammu,
Jammu 180001, India} \affiliation{Joint Institute for Nuclear
Research, Dubna, 141 980, Russia} \affiliation{Kent State
University, Kent, Ohio 44242, USA} \affiliation{University of
Kentucky, Lexington, Kentucky, 40506-0055, USA}
\affiliation{Institute of Modern Physics, Lanzhou, China}
\affiliation{Lawrence Berkeley National Laboratory, Berkeley,
California 94720, USA} \affiliation{Massachusetts Institute of
Technology, Cambridge, MA 02139-4307, USA}
\affiliation{Max-Planck-Institut f\"ur Physik, Munich, Germany}
\affiliation{Michigan State University, East Lansing, Michigan
48824, USA} \affiliation{Moscow Engineering Physics Institute,
Moscow Russia} \affiliation{City College of New York, New York City,
New York 10031, USA} \affiliation{NIKHEF and Utrecht University,
Amsterdam, The Netherlands} \affiliation{Ohio State University,
Columbus, Ohio 43210, USA} \affiliation{Old Dominion University,
Norfolk, VA, 23529, USA} \affiliation{Panjab University, Chandigarh
160014, India} \affiliation{Pennsylvania State University,
University Park, Pennsylvania 16802, USA} \affiliation{Institute of
High Energy Physics, Protvino, Russia} \affiliation{Purdue
University, West Lafayette, Indiana 47907, USA} \affiliation{Pusan
National University, Pusan, Republic of Korea}
\affiliation{University of Rajasthan, Jaipur 302004, India}
\affiliation{Rice University, Houston, Texas 77251, USA}
\affiliation{Universidade de Sao Paulo, Sao Paulo, Brazil}
\affiliation{University of Science \& Technology of China, Hefei
230026, China} \affiliation{Shandong University, Jinan, Shandong
250100, China} \affiliation{Shanghai Institute of Applied Physics,
Shanghai 201800, China} \affiliation{SUBATECH, Nantes, France}
\affiliation{Texas A\&M University, College Station, Texas 77843,
USA} \affiliation{University of Texas, Austin, Texas 78712, USA}
\affiliation{Tsinghua University, Beijing 100084, China}
\affiliation{United States Naval Academy, Annapolis, MD 21402, USA}
\affiliation{Valparaiso University, Valparaiso, Indiana 46383, USA}
\affiliation{Variable Energy Cyclotron Centre, Kolkata 700064,
India} \affiliation{Warsaw University of Technology, Warsaw, Poland}
\affiliation{University of Washington, Seattle, Washington 98195,
USA} \affiliation{Wayne State University, Detroit, Michigan 48201,
USA} \affiliation{Institute of Particle Physics, CCNU (HZNU), Wuhan
430079, China} \affiliation{Yale University, New Haven, Connecticut
06520, USA} \affiliation{University of Zagreb, Zagreb, HR-10002,
Croatia}

\author{B.~I.~Abelev}\affiliation{University of Illinois at Chicago, Chicago, Illinois 60607, USA}
\author{M.~M.~Aggarwal}\affiliation{Panjab University, Chandigarh 160014, India}
\author{Z.~Ahammed}\affiliation{Variable Energy Cyclotron Centre, Kolkata 700064, India}
\author{A.~V.~Alakhverdyants}\affiliation{Joint Institute for Nuclear Research, Dubna, 141 980, Russia}
\author{B.~D.~Anderson}\affiliation{Kent State University, Kent, Ohio 44242, USA}
\author{D.~Arkhipkin}\affiliation{Brookhaven National Laboratory, Upton, New York 11973, USA}
\author{G.~S.~Averichev}\affiliation{Joint Institute for Nuclear Research, Dubna, 141 980, Russia}
\author{J.~Balewski}\affiliation{Massachusetts Institute of Technology, Cambridge, MA 02139-4307, USA}
\author{L.~S.~Barnby}\affiliation{University of Birmingham, Birmingham, United Kingdom}
\author{S.~Baumgart}\affiliation{Yale University, New Haven, Connecticut 06520, USA}
% ...

\collaboration{STAR Collaboration}\noaffiliation

\date{\today} 

\begin{abstract}
We report on a measurement of the
$\Upsilon\textrm{(1S+2S+3S)}\rightarrow e^+e^-$ cross section at
midrapidity in \pp\ collisions at $\sqrt{s}=200$ GeV.  We find the
cross section to be $114\pm 38~\textrm{(stat. + fit)}
^{+23}_{-24}~\textrm{(syst.)}$ pb. Perturbative QCD calculations at
next-to-leading order in the Color Evaporation Model are in
agreement with our measurement, while calculations in the Color
Singlet Model underestimate it by 2$\sigma$. Our result is
consistent with the trend seen in world data as a function of the
center-of-mass energy of the collision and extends the availability
of \upsi\ data to RHIC energies. The dielectron continuum in the
invariant mass range near the \upsi\ is also studied to obtain a
combined cross section of Drell-Yan plus \bbbar$\ \rightarrow
\epluseminus$.
\end{abstract}
\keywords{Upsilon production, Quarkonium, Bottomonium}

\pacs{13.20.Gd, 14.40.Pq, 13.75.Cs, 12.38.-t, 12.38.Mh, 25.75.-q,
25.75.Nq, 25.75.Cj} \maketitle

\section{Introduction}
The main focus of the heavy flavor program at RHIC is to investigate
the properties of the Quark-Gluon Plasma by studying its effect on
open heavy flavor and quarkonia production.  $J/\psi$ suppression
induced by Debye screening of the static Quantum Chromo Dynamics
(QCD) potential between $c\bar{c}$ pairs was originally hailed as an
unambiguous signature of Quark-Gluon Plasma (QGP)
formation~\cite{matsui}. However, this simple picture is complicated
by competing effects that either reduce the yield, such as co-mover
absorption~\cite{gavin, blaizot}, or enhance it, such as in
recombination models~\cite{grandchamp, thews, andronic}. Recently, a
growing interest in studying the \upsi\ meson and its excited states
has been kindled as it is expected that color screening will be the
dominant effect contributing to any observed suppression of
bottomonium production in heavy-ion collisions. A full spectroscopy
of quarkonia states is now clearly recognized as one of the key
measurements needed to understand the matter produced in high-energy
heavy-ion collisions~\cite{Digal:2001ue}. In particular, it has been
recognized that data on the particle spectra of bottomonia can
provide valuable information to constrain QGP
models~\cite{Gunion:1996qc}. Due to the low production cross section
of \bbbar\ at RHIC ($\sigma_{b\textrm{-}\bar{b}}\approx$1.9 $\mu$b,
\cite{Cacciari:2005rk}), recombination effects in $A+A$ collisions
are negligible. At the same time, the interaction cross section of
bottomonium with the abundantly produced hadrons in these collisions
is small~\cite{Lin:2000ke}, so suppression due to absorption by
hadronic co-movers is expected by these models to be relatively
unimportant. However, it will still be important to study \upsi\
production in \dAu\ collisions since available measurements by
E772~\cite{Alde:1991sw} of cold nuclear matter effects on \upsi\
production at lower energy show some suppression. Nevertheless, the
amount of suppression seen for the \upsi\ family is measured to be
smaller than for charmonia. Therefore, bottomonium is expected to be
a cleaner probe of high-temperature color screening effects.

In addition to its important role in establishing deconfinement, a
measurement of the \upsi\ 1S, 2S, and 3S states in \pp\ and
heavy-ion collisions can help to set limits on the medium
temperature. The quarkonium measurements help in reaching these key
goals because ($i$) an observation of suppression of \upsi\
production in heavy-ions relative to \pp\ would be a strong argument
in support of Debye screening and therefore of
deconfinement~\cite{Mocsy:2007jz}, and ($ii$) the sequential
suppression pattern of the excited states is sensitive to the
temperature reached in the medium~\cite{Digal:2001ue}. In this
regard, lattice QCD studies have seen a burst of activity in recent
years. Studies of quarkonia spectral functions and potential models
based on lattice QCD indicate that while the \upsithree\ melts even
before the deconfinement transition and the \upsitwo\ is likely to
melt at RHIC (\sqrts = 200 \gev), the \upsione\ is expected to
survive~\cite{Digal:2001ue,Digal:2001iu,wong}. Recent results
\cite{Mocsy:2007jz,Mocsy:2007yj} indicate further that almost all
quarkonia states (\jpsi, $\psi'$, $\chi_c$, $\chi_b$, \upsitwo) melt
below 1.3 \Tc\ and the only one to survive to higher temperature is
the \upsione, which melts at 2\Tc, where $\Tc \approx 175$ MeV is
the critical temperature for the parton-hadron phase transition.
Therefore, a systematic study of all quarkonia states in \pp, \dAu,
and \AuAu\ collisions will provide a clearer understanding of the
properties of the Quark-Gluon Plasma.


Suppression of the \upsitwo\ and \upsithree\ should be measurable at
RHIC energies with increased integrated luminosity. In the near
future, the larger luminosities proposed by the RHIC II
program~\cite{rhic2} will allow for a statistically significant
measurement of all 3 states. With the objective of embarking in such
a long program, one of the first steps is to establish a baseline
cross section measurement of the bottomonia states in \pp\
collisions. There are no previous measurements of \upsi\ production
in \pp\ at the top RHIC energy for heavy ions (an upper limit was
estimated in the 2004 data with only half of the
calorimeter~\cite{kollegger}). The luminosities available at RHIC in
the 2006 run provided the first opportunity to measure bottomonium
at the previously unexplored center-of-mass energy of $\sqrts=200$
\gev.  A dedicated trigger algorithm exploiting the capabilities of
the STAR electromagnetic calorimeter is essential for this
measurement, and its development in STAR allows the \upsi\ family to
be studied in the \epluseminus\ decay channel. In this paper, we
report our result for the \upsi(1S+2S+3S) cross section at
midrapidity, obtained with the STAR detector in \pp\ collisions at
$\sqrt{s}=200$ GeV via the \epluseminus\ decay channel.  This
measurement uses an integrated luminosity of 7.9 pb$^{-1}$ collected
during RHIC Run VI (2006). We compare our data to perturbative QCD
calculations done at next-to-leading order (NLO pQCD) in the Color
Evaporation Model (CEM)~\cite{rhic2} and in the Color Singlet Model
(CSM)~\cite{Brodsky:2009cf}.

The article is organized as follows. Section~\ref{sec:DetOverview}
explains the detector setup. The details of the quarkonia triggers
are explained in Sec.~\ref{sec:QuarkoniaTriggers}. The acceptance
and trigger efficiency are discussed in Sec.~\ref{sec:AccTrigEff}.
After a detailed discussion of the data analysis procedure in
Section~\ref{sec:OfflineAnalysis}, we present our results and
compare with pQCD calculations and with available data in
Sec.~\ref{sec:ResultsDiscussion}. Our findings are summarized in
Sec.~\ref{sec:Conclusions}.

\section{Detector Overview}\label{sec:DetOverview}
The main detectors used in the STAR quarkonia program are the Time
Projection Chamber (TPC)~\cite{tpc}, the Barrel Electromagnetic
Calorimeter (BEMC)~\cite{emc}, and the Beam-Beam Counters
(BBC)~\cite{bbc}.

The BBCs are segmented scintillator rings covering the region $3.3 <
|\eta| < 5.0$. The STAR \pp\ minimum-bias trigger requires the
coincidence of hits in the BBCs on opposite sides of the interaction
region, and is used to monitor the integrated luminosity for
measuring absolute cross sections.  The BBC acceptance and
efficiency were studied previously~\cite{Adams:2003kv}, and it was
found from simulations that for \pp\ non--singly diffractive (NSD)
events the BBC trigger was $87\pm8$\% efficient. The absolute \pp\
cross section seen by the BBC trigger was measured via a van der
Meer scan to be 26.1 $\pm$ 0.2 (stat.) $\pm$ 1.8 (syst.) mb.

The TPC and BEMC have a large acceptance at midrapidity: the single
track coverage of the TPC+BEMC in pseudorapidity is $|\eta|<1$ and
complete in azimuth. The BEMC is divided in $\eta$ and $\varphi$
into 4800 towers. The size of each tower in $\eta\times\varphi$ is
$0.05 \times 0.05$ rad. The geometrical acceptance of the STAR BEMC
for detecting both electrons from an \upsi\ decay is about
$\approx30\%$ over all phase space, and $\approx60\%$ for \upsi's in
the kinematic range $|y|<0.5$ and $\pT<5$ \gevc. We focus on this
kinematic range, where the acceptance of the detector is optimal.

The capability of the TPC+BEMC for electron identification and the
triggering capabilities of the BEMC are the two pillars of the STAR
quarkonium program. In particular, the BEMC trigger allows us to
sample the full luminosity delivered by RHIC to look for the
high-mass dielectron signals characteristic of the
$\Upsilon\rightarrow e^+e^-$ decay.

\section{The STAR Quarkonia Triggers}\label{sec:QuarkoniaTriggers}
The STAR quarkonia trigger is a two-stage system comprising a
Level-0 (L0) hardware component~\cite{Bieser:2002ah} (decision time
of $\approx 1~\mu$s) and a Level-2 (L2) software component (decision
time of $\approx 100~\mu$s in \pp, and $\approx 400~\mu$s in \AuAu).
There were two separate \upsi\ triggers used during the 2006 run; in
the following we refer to them as Trigger I and Trigger II. They
were identical in concept, with the only difference being an
increase in thresholds, which was necessary to reduce the trigger
rate and reduce deadtime while data-taking. We discuss the triggers
next, and list all the parameters and thresholds. These were chosen
based on simulations and on expected trigger rates based on
calorimeter data taken in 2003.


\subsection{\upsi\ L0 Trigger}
The L0 trigger is a minimum-bias \pp\ trigger with the additional
requirement of signals in the BEMC consistent with a high energy
electron. The energy deposited in the calorimeter towers is measured
by collecting scintillation light from the electromagnetic shower.
The signal is digitized, pedestal subtracted, and sent to the L0
Data Storage and Manipulation (DSM) hardware as a 6-bit ADC value
for triggering. The calorimeter calibration is done such that tower
ADC values are proportional to deposited electromagnetic transverse
energy (\ET), making it useful for triggering on high-\pT\
electrons. The L0 trigger decision is made for every bunch crossing.

For the \upsi\ analysis, the L0 decision was based on two
quantities. The first is related to the signal in distinct towers.
The tower with the highest \ET, as represented by the DSM-ADC value,
is called the ``High Tower'' (HT) in the event. The second quantity
is a sum of towers in fixed $\eta$-$\varphi$ regions of the BEMC,
where each region comprises 4$\times$4 towers. Each of these regions
is named a ``Trigger Patch'' (TP). Since each tower covers 0.05
units in $\eta$ and 0.05 radians in $\varphi$, the TP coverage is
$0.2\times0.2$ rad. The L0 trigger is issued for Trigger I (Trigger
II) if all of the following three conditions are met: ($i$) an event
has a High Tower with a DSM-ADC value $>$ 12 (16), corresponding to
\ET $\approx$ 2.6 (3.5) \gev\ deposited in the tower; ($ii$) the
Trigger Patch containing the High Tower has a total DSM-ADC sum over
the 16 towers in the patch with a value $>$ 17 (19), corresponding
to \ET\ $\approx$ 3.8 (4.3) \gev, and ($iii$) the STAR \pp\
minimum-bias trigger is met. The minimum-bias trigger is based on a
BBC coincidence and is described elsewhere~\cite{Adams:2003kv}.  The
coincidence of High-Tower and Trigger Patch triggers will be
referred to as ``HTTP'' in the remainder of the paper.

\begin{figure}[t]
\includegraphics*[width=\columnwidth]{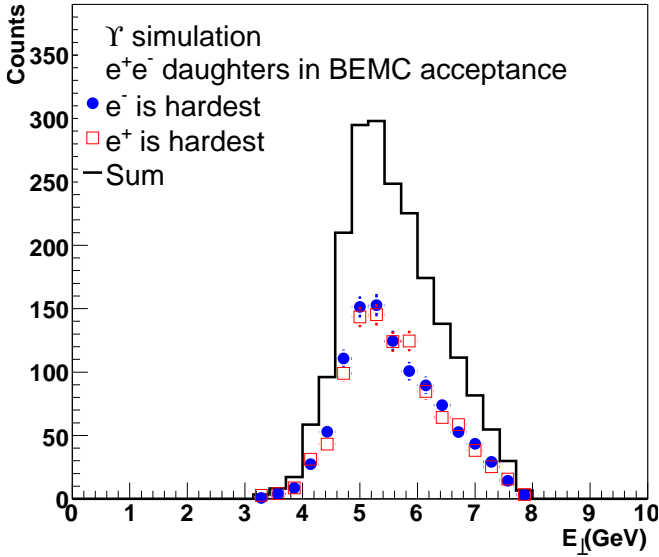}
\caption{\label{fig:UpsElePtSpec} The \ET\ distribution for
simulated \upsi\ daughter electron (filled circle) or positron (open
square) with the highest \ET. We show only daughters that fall in
the STAR BEMC geometrical acceptance. The histogram is the sum of
the two distributions. The L0 HT Trigger II threshold of 16 counts
corresponds to $\ET\approx 3.5$ \gev.}
\end{figure}

\begin{figure}[t]
\includegraphics*[width=\columnwidth]{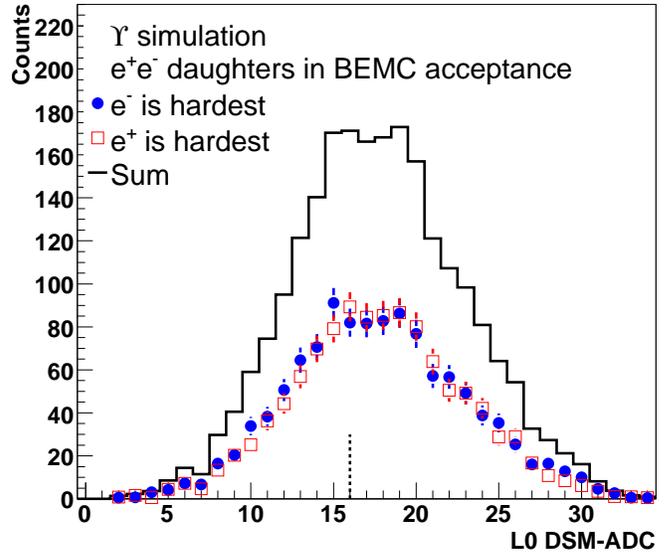}
\caption{\label{fig:UpsEleAdcSpec} The L0 DSM-ADC distribution for
the electron (filled circle) or positron (open square) with the
highest ADC.  The histogram is the sum. The simulated \upsi\ is
triggered when one of its daughter electrons is above the dashed
line, which indicates the L0 HT Trigger II threshold of 16 counts.}
\end{figure}

Figure~\ref{fig:UpsElePtSpec} shows the \pT\ distribution for
simulated electrons and positrons from \upsi\ decays in which both
daughters are within the STAR acceptance.  For each decay, we plot
the \pT\ distribution of the electron (filled circles) or positron
(open squares) with the highest \pT (the abscissa is simulated \pT).
The electron and positron distributions are identical, as expected
for a two-body decay into daughters with identical masses. The
histogram in the figure is the sum, \ie, the \pT\ distribution for
the hardest of the two daughters. The corresponding DSM-ADC values
used in the L0 trigger (which are proportional to \pT) are shown in
Fig.~\ref{fig:UpsEleAdcSpec}. The L0 HT Trigger II threshold of 16
is shown by the vertical dashed line.

From the \pT\ and ADC distributions for the hardest \upsi\ daughter
in Figs.~\ref{fig:UpsElePtSpec} and \ref{fig:UpsEleAdcSpec}, the
average values are $\langle\pT\rangle =$ 5.6 GeV/c and $\langle
\mathrm{ADC}\rangle =$ 18.0 counts.  Since the trigger threshold is
placed at 16 ADC counts, the hardest daughter will typically fire
the trigger. We find that 25\% of the \upsi's produced at
midrapidity have both daughters in the BEMC acceptance and at least
one of them can fire the L0 trigger. The details of the HTTP trigger
efficiency and acceptance are discussed in Sec.
~\ref{sec:AccTrigEff}.

\subsection{\upsi\ L2 Trigger}
Once a L0 HTTP trigger is issued, the information for all detectors
in STAR begins to be digitized.  During this time, the L2 system can
use the information from the 4800 individual calorimeter towers to
decide whether to keep the event or to abort the readout. The L2
system can use the full energy resolution of the calorimeter (a
10-bit ADC value, in contrast to the 6-bit ADC used at L0). The L2
\upsi\ trigger makes the decision by looking for calorimeter
signatures consistent with the production of a high invariant mass
electron-positron pair. (Since the L2 trigger uses BEMC information
only and does not distinguish electrons from positrons, we will
refer to electrons and positrons simply as ``electrons'' in all
discussions of the L2 trigger.) At a minimum, one pair of candidate
electrons is required. They are processed as follows. (We will
denote kinematical quantities obtained in the L2 stage with the
superscript ``L2'' to distinguish them from similar quantities
obtained during the offline analysis.)

The algorithm starts by searching for all towers above the L0 High
Tower threshold. Each of these is treated as a seed for a 3-tower
cluster. To produce the clusters, we search in the $\eta$-$\varphi$
region around each High Tower, with a search window of $3\times3$
towers.  This area is smaller than the TP size in order to focus on
electron finding, as our simulations show that electrons will likely
have most of their energy contained in only 3 towers. We sort these
8 surrounding towers according to their measured energy and pick the
two highest-\ET\ towers in this list to produce a 3-tower cluster.
This L2 tower clustering gives a better estimate of the electron
energy compared to a single tower due to possible shower leakage
into neighboring towers. If the energy of this 3-tower cluster is
greater than 4.0 \gev\ it is considered for further processing, and
we label such clusters as ``L2 Cluster-1''.

We next look for additional electron candidates in the event. While
it is possible for an event to have two towers above the HT
threshold, the majority of events will have only one. We select
additional electron candidates by starting with towers which have
$\ET\gtrsim0.7$ \gev\ (10-bit ADC of at least 75 counts after
pedestal subtraction). Starting from these second seeds, we again
construct 3-tower clusters (L2 Cluster-2) using the procedure
outlined above. We require that the L2 Cluster-2 energy be
$E_2^{L2}> 2.5$ \gev. After the complete iteration to find electron
clusters, we make all possible cluster pairs (where each pair must
have at least one L2 Cluster-1) and calculate two pairwise
quantities: the opening angle and the invariant mass of the pairs.
These are calculated under the approximation that the vertex
location is in the center of the detector and that the electrons
travel in a straight line. Since the majority of the $\epluseminus$
pairs from \upsi\ decays have a large opening angle ($\theta$), we
look for pairs with $\cos(\theta_{12}^{L2})<0.5$,
where $\theta_{12}^{L2}$ is the angle between the two clusters
calculated in the L2 algorithm. With the combined information on the
energy of the two clusters and their opening angle, we reconstruct
the invariant mass via the approximate formula
$M_{12}^{L2}\approx\sqrt{2E_1^{L2}E_2^{L2}(1-\cos\theta_{12}^{L2})}$.
We can neglect the electron mass of 0.511 \mevcc\ as it would only
contribute $\approx 1$ \mevcc, which is small for ultra-relativistic
electrons. For comparison, the straight-line approximation and the
energy resolution result in a mass resolution of $\approx 900$
\mevcc\ as will be discussed in the following section. We select
events with cluster-pair invariant masses in the range $6 <
M_{12}^{L2} < 15$ \gevcc.  If there is any pair that satisfies both
of the pairwise conditions in the event, the algorithm issues an L2
trigger.

\subsection{Trigger Performance}
\label{sec:TriggerPerformance} In order to evaluate the trigger
performance, we applied the Trigger II cuts to the Trigger I data.
To do this, the exact same trigger condition was applied offline on
the recorded values of the original trigger input data. Since events
satisfying Trigger II cuts also satisfy Trigger I cuts, applying the
tighter set of Trigger II cuts to the Trigger I data offline allows
us to make a single dataset with uniform properties for the entire
2006 run. The integrated luminosity for Trigger I was
$\mathcal{L}=3.12\pm 0.22 $ (syst.) \invpb, and for Trigger II it
was $\mathcal{L}= 4.76\pm 0.33 $ (syst.) \invpb, giving a total
integrated luminosity of 7.9 $\pm$ 0.6 \invpb, where the 7\%
uncertainty originates from the uncertainty in the BBC measurement
of the cross section as determined by a van der Meer
scan~\cite{Adams:2003kv}.
\begin{figure}[t]
\includegraphics*[width=\columnwidth]{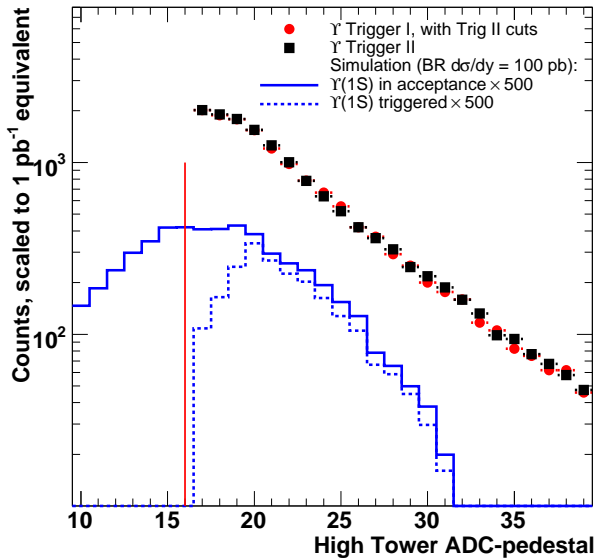}
\caption{\label{fig:HighTowerAdc} The L0 DSM-ADC distribution
($\propto \ET$) for the highest EMC tower of a candidate pair. We
show Trigger I data after applying Trigger II thresholds (red
circles) and Trigger II data (black squares). The yields are
normalized by the integrated luminosity. The histograms are from
simulation of \upsione, showing the corresponding distribution for
electron daughters satisfying acceptance (solid line) and trigger
(dotted line) requirements. The simulation histograms are normalized
assuming $\mathcal{B}\times d\sigma/dy$= 100 pb, times a factor of
500 for clarity. The vertical line is the Trigger II threshold of 16
counts. The histograms have the bin centers set at integer values to
match the integer nature of ADC counts.}
\end{figure}

\begin{figure}[t]
\includegraphics*[width=\columnwidth]{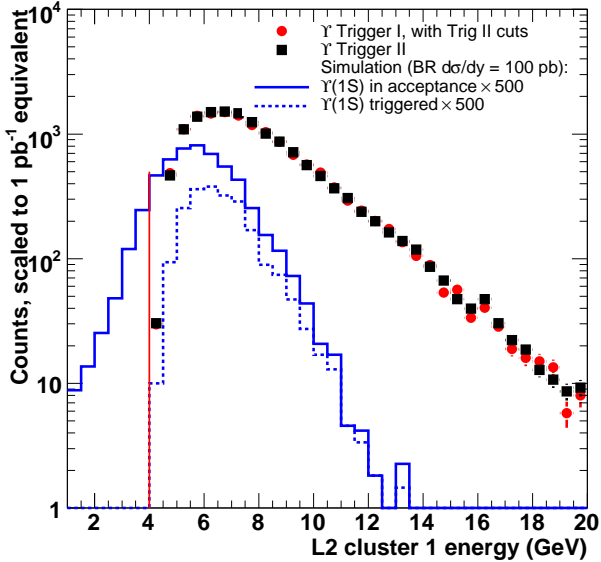}
\caption{\label{fig:ClusterE} The distribution of L2 Cluster-1
energy $E_1^{L2}$
 for all towers above the HT threshold in a trigger
patch above the TP threshold. The L2 trigger requires $E_1^{L2}>
4.0$ \gev\ (vertical line).  The line histograms show the \upsi\
(from simulation) after acceptance requirements (solid line), and
after all trigger requirements (dashed line). The normalization and
scaling factors are the same as in Fig.~\ref{fig:HighTowerAdc}.}
\end{figure}

\begin{figure}[t]
\includegraphics*[width=\columnwidth]{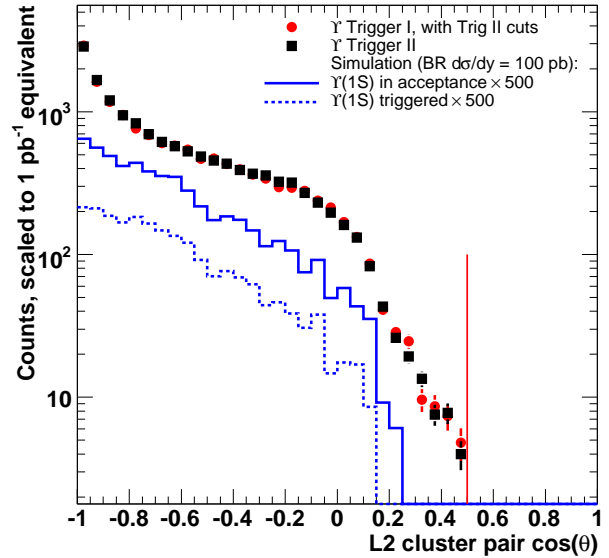}
\caption{\label{fig:L2CosTheta} The distribution of cosine of L2
opening angle $\theta_{12}^{L2}$  for accepted events. The line
histograms show the \upsi\ distribution after acceptance
requirements (solid line), and after all trigger requirements
(dashed line). The vertical line shows the location of the trigger
threshold. Normalization and scaling factors are the same as in
Fig.~\ref{fig:HighTowerAdc}.}
\end{figure}

\begin{figure}[t]
\includegraphics*[width=\columnwidth]{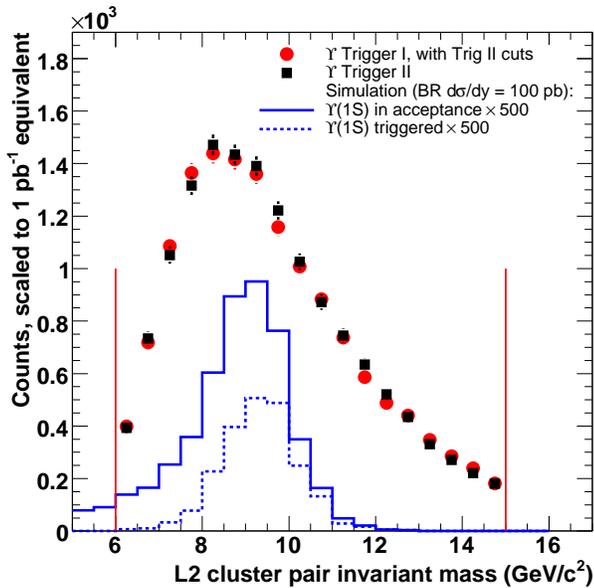}
\caption{\label{fig:L2InvM} The L2 invariant mass $M_{12}^{L2}$
distribution for accepted events. The line histograms show the
\upsi\ $M_{12}^{L2}$ distribution after acceptance requirements
(solid line), and after all trigger requirements (L0 HTTP trigger,
L2 cluster, L2 opening angle and L2 invariant mass) of the
\epluseminus\ daughters (dashed line).  Vertical lines show location
of the trigger thresholds. Normalization and scaling factors are the
same as in Fig.~\ref{fig:HighTowerAdc}.}
\end{figure}

Figure~\ref{fig:HighTowerAdc} shows the ADC distribution of the
tower with the largest \ET\ for each candidate pair seen at the
trigger level in an event (in very few cases we find more than one
tower above the L0 threshold, so most events have only one candidate
pair). The two triggered data sets are shown, with Trigger II
displayed as squares and Trigger I (after applying Trigger II cuts)
displayed as circles.  The rejection factor achieved with Trigger
II, defined as the number of minimum bias events counted by the
trigger scalers divided by the number events where the trigger was
issued, was found to be $1.8\times 10^5$. The distributions seen at
the trigger level only include information from the BEMC.  This
causes the trigger rate seen in the experiment to be dominated by
background from di-jet events with two nearly back-to-back $\pi^0$s.
This di-jet background is removed offline when including tracking
information from the TPC. The trigger distributions are scaled by
the overall luminosity in order to compare the relative
normalization of the two datasets. The scale is chosen such that the
counts in each triggered dataset correspond to an integrated
luminosity of 1 \invpb. The relative luminosity normalization
between the datasets agrees to a level of $\approx1\%$. The
solid-line histogram is from simulated \upsione\ after acceptance
selection, requiring both electrons to deposit at least 1 GeV of
energy in the BEMC. It shows the spectrum for the daughter with the
highest ADC count. The spectrum for the simulated \upsione\ events
that satisfy all the trigger requirements is shown as the
dashed-line histogram. The vertical line at 16 ADC counts represents
the L0 ADC threshold for Trigger II.  In order to compare the size
of the expected \upsi\ signal relative to the trigger background
from di-jets, the histograms from the simulations for the \upsione\
are scaled with two factors. The first factor corresponds to
normalizing to an integrated luminosity of 1 \invpb\ assuming a
cross section times branching ratio into \epluseminus\ for the
\upsione\ at midrapidity of 100 pb, chosen because it is of the
expected order of magnitude. From this, we expect one upsilon every
$\approx3000$ triggers, so we use a second multiplicative factor of
500 for display purposes. The spectral shape of the data in
Fig.~\ref{fig:HighTowerAdc} is therefore dominated by di-jet
background.  The same normalization and scale factors are used in
Figs.~\ref{fig:ClusterE}--\ref{fig:L2InvM} for comparing the
trigger-level background distributions and the expected \upsione\
signal.

Figure~\ref{fig:ClusterE} shows the distribution of the L2 Cluster-1
energy for all clusters that include a tower above the HT DSM-ADC
threshold of 16 ADC and that have their corresponding Trigger Patch
DSM-ADC sum above the threshold of 19 ADC counts. It can be seen
that the peak of the L2-cluster distribution near 6 \gev\ is not
right at the threshold of 4 \gev\ (vertical line). This must happen
because another trigger selection that is correlated to the
measurement of the L2-cluster is being applied. In our case it is
the L0 selection, which consists of both the HT and the TP
requirement. Once we see a tower with energy above the HT threshold
that is also in a trigger patch with energy above the TP threshold,
the energy in the 3-tower cluster for that tower will likely already
be above the L2 E1 threshold.

The distribution for the L2 opening angle $\theta_{12}^{L2}$ is
shown in Fig.~\ref{fig:L2CosTheta}. It is highly peaked towards
back-to-back topologies, much more so than the distribution from
simulated \upsi's.  This again reflects the fact that the triggered
distribution is dominated by back-to-back $\pi^0$'s. The majority of
these are rejected offline when requiring the presence of a
corresponding electron track in the TPC.

The L2 invariant mass distribution shown in Fig.~\ref{fig:L2InvM} is
peaked at $\approx$ 8 \gevcc\ due to the cluster energy
requirements. The vertical lines depict the lower and upper
thresholds at 6 and 15 \gevcc. The histogram shows the simulated
\upsione\ before any trigger cuts (solid) and after passing all the
previous thresholds (dashed). The trigger preferentially rejects
lower L2-mass \upsi\ events, because these can happen when the
energy clusters measured in the calorimeter are lower than their
average, and these clusters are preferentially rejected by the
algorithm. From these simulations we estimate the \upsi\ mass
resolution of the L2 trigger to be $849 \pm 8$ \mevcc. It is
dominated by the BEMC energy resolution and by the straight-line
approximation used to calculate the opening angle in the L2 trigger
algorithm.  This is about an order of magnitude larger than the
offline mass resolution, where the electron trajectories and momenta
are obtained from tracking. Since all the \upsi's that reach this
stage are contained within the invariant-mass limits, this invariant
mass cut serves mainly to reject background.

In addition, the L2 information (such as that shown in
Figs.~\ref{fig:ClusterE}--\ref{fig:L2InvM}) is available online
after every run.  These distributions serve as useful diagnostic
tools during data taking.

\section{\upsi\ Acceptance and Trigger Efficiency}
\label{sec:AccTrigEff}

To determine the geometrical acceptance of the detector for
measuring \upsi\ $\rightarrow$ \epluseminus, we combined two types
of events in the following way. For the first type, we performed
GEANT simulations of the \upsi\ $\rightarrow$ \epluseminus\ decays
and propagated them through the detector geometry. The simulations
were done with uniform population of the $\pT$-$y$ phase space,
folded with a Gaussian in $y$ ($\sigma=1$) and a realistic $\pT$
distribution. We chose the form $dN/d\pT\propto \pT/(\exp(\pT/T)+1)$
with the parameter $T=2.2$ \gevc\ obtained from a fit to CDF
data~\cite{Acosta:2001gv}. The dependence of the acceptance on the
choice of $\pT$ distribution is negligible. This was verified by
using functional forms derived from data at lower energy. Each
simulated decay was combined with a simulated \pp\ minimum-bias
event using the PYTHIA event generator~\cite{Sjostrand:2008vc} with
CDF Tune A settings~\cite{Field:2005sa}. For the second type, we
used a set of data collected by triggering on a RHIC bunch crossing.
These were labeled ``zero-bias'' as they do not require signals in
any of the STAR detectors. These zero-bias events do not always have
a collision in the given bunch crossing, but given that in the 2006
\pp\ run the dominant contribution to the TPC occupancy was from
pileup, these events will include the pileup from out-of-time
collisions and all detector effects. When the zero-bias events are
combined with simulated events, they provide the most realistic
environment to study the detector efficiency and acceptance. Each
PYTHIA+\upsi\ event was embedded into a zero-bias event.  The vertex
position chosen for the simulated event was sampled from a realistic
distribution of event vertex positions obtained from the \upsi\
triggered data. With this procedure, we estimate both the trigger
efficiency and the reconstruction efficiency (discussed in
Sec.~\ref{sec:OfflineAnalysis}).

\begin{figure}[t]
\includegraphics*[width=\linewidth]{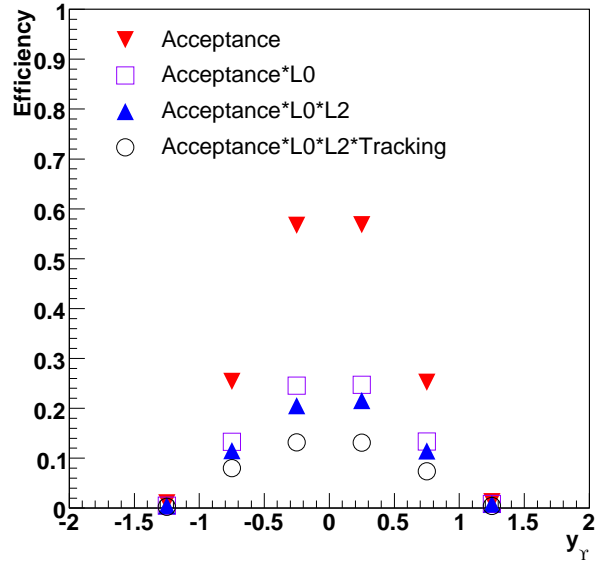}
\caption{\label{fig:RapidityEffAcc} Combined acceptance $\times$
efficiency as a function of rapidity $y_\Upsilon$ for the decay
\upsi $\rightarrow$ \epluseminus\ in STAR. Down triangles: BEMC
acceptance only; open squares: adding L0 requirement; upright
triangles: adding L2 requirement; open circles: adding TPC
acceptance+TPC tracking cuts and TPC-BEMC matching.}
\end{figure}

For our purposes, a simulated \upsi\ is considered to fall in the
acceptance of the detector if each of its decay electrons deposits
at least 1 \gev\ in a BEMC tower, as simulated by GEANT. We find
that the acceptance depends strongly on the \upsi\ rapidity
($y_{\Upsilon}$), but weakly on its transverse momentum
($\pT_{\Upsilon}$).  At $|y_{\Upsilon}|< 0.5$ the acceptance is
57\%, dropping to below 30\% beyond $|y_{\Upsilon}| = 0.5$, and is
close to zero beyond $|y_{\Upsilon}| = 1.0$. This is illustrated in
Fig.~\ref{fig:RapidityEffAcc}, where the downward triangles depict
the geometrical acceptance of the BEMC detector
$\epsilon_{\mathrm{geo}}$ obtained from the analysis of the
simulated \upsi\ decays in the real data events.

For those simulated \upsi's that are accepted, we calculate the
trigger efficiencies.  This requires simulations of the BEMC
response, digitization, and running the result through the offline
reconstruction software chain.  The HTTP requirement and the L2
trigger condition cut out an additional fraction of the events, and
the combined acceptance and trigger efficiency is shown in
Fig.~\ref{fig:RapidityEffAcc}. The squares include the effect of the
HTTP trigger requirement $\epsilon_{\textrm{L0}}$ and the upright
triangles include the L2 trigger condition $\epsilon_{\textrm{L2}}$.
The open circles include TPC acceptance plus tracking efficiency
$\epsilon_{\textrm{track}}$ and TPC-BEMC matching $\epsilon_{R}$.
These selections, as well as the particle identification cuts, are
discussed in Sec.~\ref{sec:OfflineAnalysis}. For the region
$|y_{\Upsilon}|<0.5$, the acceptance and efficiencies are given in
Table~\ref{tab:triggerEff}.

\begin{table}
\caption{\label{tab:triggerEff} Acceptance, trigger efficiency, and
tracking efficiency for reconstructing $\upsi \rightarrow
\epluseminus$ in STAR in the kinematic region $|y_{\Upsilon}|<0.5$.
The first 4 rows are for the 1S and the last row is for the cross
section--weighted 1S+2S+3S combination. All systematic uncertainties
are listed in Table~\ref{tab:SystUncert}.}
\begin{ruledtabular}
\begin{tabular}{lc}
    Quantity                & \textrm{Value}   \\
    \hline
  $\epsilon_{\mathrm{geo}}$ & 0.57    \\
  $\epsilon_{\mathrm{geo}}\times\epsilon_{\textrm{L0}}$ & 0.25 \\
  $\epsilon_{\mathrm{geo}}\times\epsilon_{\textrm{L0}}\times\epsilon_{\textrm{L2}}$ & 0.21 \\
  $\epsilon_{\mathrm{geo}}\times\epsilon_{\textrm{L0}}\times\epsilon_{\textrm{L2}}\times\epsilon_{\mathrm{track}}\times\epsilon_{R}$ & 0.13 \\
  $\epsilon_{\mathrm{geo}}\times\epsilon_{\textrm{L0}}\times\epsilon_{\textrm{L2}}\times\epsilon_{\mathrm{track}}\times\epsilon_{R}$ (1S+2S+3S)& 0.14 \\
\end{tabular}
\end{ruledtabular}
\end{table}

The relative systematic uncertainty on the L0 HT trigger efficiency
was estimated to be $^{+5.9\%}_{-7.5\%}$ and is the dominant source
of systematic uncertainty of the \upsi\ trigger efficiency. The
systematic uncertainties for the L0 efficiency are asymmetric
because the underlying \ET\ distribution of the daughter electrons
is not flat in the threshold region (see
Fig.~\ref{fig:HighTowerAdc}). A shift to a higher threshold reduces
the yield more than a shift to a lower threshold increases it.  The
L0 HTTP trigger has largest effect on the efficiency, as only 43\%
of the events in the acceptance remain after the trigger condition
is applied. After including the L2 trigger, tracking, and TPC-BEMC
matching efficiencies, we obtain a combined efficiency of 13.2\% for
the \upsione.

A similar procedure was applied to the \upsitwo\ and \upsithree\
states, which have slightly higher efficiencies due to their larger
masses. We calculated a weighted average among the 3 states,
including the branching ratio and the ratio of cross sections, in
order to obtain an average efficiency to be applied to the measured
yield of \upsi(1S+2S+3S). For this we use the branching ratios
compiled by the PDG~\cite{pdg}, and cross section ratios from NLO
pQCD calculations in the Color Evaporation Model (CEM). The
calculation used a bottom quark mass of $m_{b}=4.75$ \gevcc, the
PDFs used are MRST HO~\cite{Martin:1998sq}, the choice of scale is
$\mu=m_T$, and the center-of-mass energy is $\sqrt{s}=200$
GeV~\cite{rhic2}. The branching ratios and cross sections we used
for this purpose are shown in
Table~\ref{tab:BranchRatiosAndCrossSec}.
\begin{table}[tb]
\caption{\label{tab:BranchRatiosAndCrossSec} Branching fractions for
$\Upsilon(\textrm{nS})\rightarrow e^+e^-$~\cite{pdg} and total cross
sections at $\sqrt{s}=200$ GeV from an NLO CEM model~\cite{rhic2}.
See text for details.}
\begin{ruledtabular}
\begin{tabular}{lcr}
$\Upsilon$ state & $\mathcal{B}$ (\%) & $\sigma$ (nb)\\
\hline
$\upsione$ & 2.38$\pm$0.11 & 6.60\\
$\upsitwo$ & 1.91$\pm$0.16 & 2.18\\
$\upsithree$ & 2.18$\pm$0.21 & 1.32\\
\end{tabular}
\end{ruledtabular}
\end{table}

This procedure results in a combined acceptance, trigger efficiency,
track finding efficiency, and TPC-BEMC matching efficiency of 14.3\%
for the averaged \upsi(1S+2S+3S) combination, see the last row of
Table~\ref{tab:triggerEff}. We estimate the sensitivity of the
efficiency to the values in Table~\ref{tab:BranchRatiosAndCrossSec}
in two ways. ($i$) By varying the branching ratio values by their
uncertainty, we find that the efficiency varies in the range
14.2--14.4\%. ($ii$) Instead of using the numbers from Table
\ref{tab:BranchRatiosAndCrossSec}, we can use measured values of the
ratios \upsitwo/\upsione\ and \upsithree/\upsione\ from
Ref.~\cite{Moreno:1990sf}.  No further input is needed because the
weighted average of the efficiency does not depend on the overall
scale of the cross section: it cancels out when averaging. By
varying these measured ratios within their uncertainty we obtain
weighted efficiencies in the range 14.1--14.4\%. Therefore, our
results do not depend strongly on the values in Table
\ref{tab:BranchRatiosAndCrossSec}.  In particular, the overall scale
of the cross section from the NLO calculations has no direct impact
on our results. Additional analysis cuts related to electron
identification are discussed in the following section.

With thresholds of $\approx 4\ \gev$, the hadron rejection power of
the BEMC towers at the energy relevant for \upsi\ decay daughters is
$e/h\approx 100$. The main source of background for the trigger
comes from high-\pT\ $\pi^0$'s decaying into two photons that
deposit energy in the BEMC. The BEMC will therefore trigger on two
high-\pT\ $\pi^0$'s with a large opening angle. These events are
typically from di-jets. Since these are also produced by
large--momentum-transfer events with low cross section, the trigger
rate is sufficiently small even in the presence of this background,
that the \upsi\ trigger can sample all collision events and is
limited only by luminosity. The triggered distributions shown in
Figs.~\ref{fig:HighTowerAdc}--\ref{fig:L2InvM} are dominated by the
$\pi^0$ photon background. These background events are rejected in
the offline analysis.

\section{$\Upsilon$ Analysis}
\label{sec:OfflineAnalysis}

During the offline analysis, we use a complete emulation of the
trigger to obtain all candidate BEMC tower pairs for an event.  We
use the TPC to select charged tracks and require that they point
close in $\eta$-$\varphi$ to the position of the candidate clusters.
The TPC also allows us to obtain improved electron kinematics
compared to those derived from BEMC information available at the
trigger level and to perform particle identification via specific
ionization. The matching of TPC tracks to BEMC clusters is also
useful for electron identification via the ratio $E/p$.

Tracks are selected based on the number of TPC points found during
the track pattern recognition and used in the fit to obtain the
track kinematic parameters.  The TPC tracks can have a maximum of up
to 45 space points. We require that all tracks have at least 52\% of
their maximum possible points, and a minimum of 20 points. The first
requirement guarantees we have no split tracks, and the second
requirement sets a floor for those tracks which have a number of
points smaller than 45 due to passing through inactive areas of the
detector. We select tracks with $\pT > 0.2$ \gevc\ to reject most
low momentum tracks in a first pass. We do not impose higher $\pT$
requirements in the track selection stage because we also require
$E/p$ matching with the BEMC, where $E>2.5$ \gev, as discussed
below. Tracks are also required to extrapolate back to a primary
vertex found in the event. We find the combined TPC acceptance times
tracking efficiency for detecting each \upsi\ daughter to be
$\epsilon_{\textrm{TPC}}=85\%$, and to be approximately independent
of electron \pT.  Hence, the pairwise efficiency can be obtained by
squaring the single particle efficiency.

\begin{figure}
\includegraphics*[width=\columnwidth]{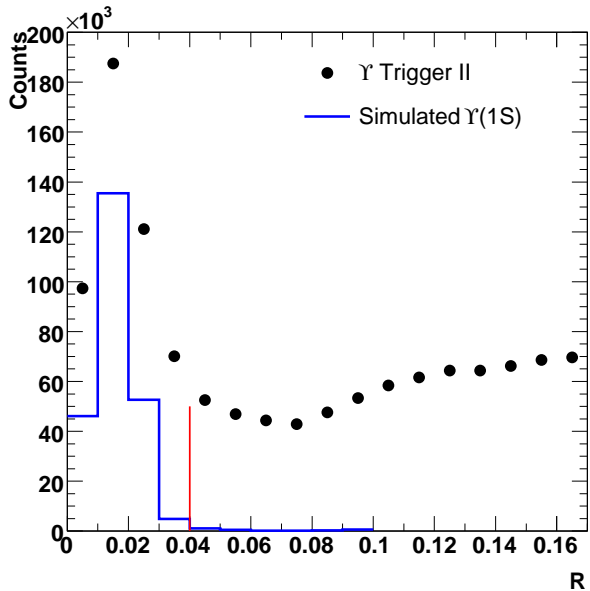}
\caption{\label{fig:Radius} The radial distance
$R=\sqrt{\Delta\eta^2+\Delta\varphi^2}$ between the TPC track and
the EMC electron cluster for simulated \upsi\ daughters (line
histogram, arbitrary scale), and the $R$ distribution from data
(circles). We reject candidate tracks with $R>0.04$ (vertical
line).}
\end{figure}

In order to guarantee that the analysis only uses tracks that could
have fired the trigger, we impose a requirement that the tracks
extrapolate close to the BEMC candidate towers. The requirement used
was for them to be within a circle of radius $R< 0.04$ in
$\eta$-$\varphi$ space of the L2 candidate clusters. We find that
this cut has an efficiency of $\epsilon_{R} = 93\%$ for a given
\upsi\ daughter electron, and has good background rejection power.
This is illustrated in Fig.~\ref{fig:Radius}, which shows the $R$
distribution for simulated \upsione\ daughters as the line
histogram. The $R$ distribution from tracks passing basic quality
cuts in the triggered events is shown as the solid circles.

An important component of the analysis is the vertex-finding
efficiency.  We find that in contrast to minimum-bias \pp\ events,
the \upsi\ triggered events have a very high vertex-finding
efficiency. The analysis benefits from the presence of the HT
trigger, since these events are likely to have a high-\pT\ track
that facilitates the task of finding the vertex.  We find that the
vertex-finding efficiency $\epsilon_{\textrm{vertex}}$ for \upsi\
events is 96.4 $\pm 0.9\%$.

In the 2006 run, the luminosity was high enough that there can be
multiple primary vertices due to pileup events in the TPC.  We find
that about $\approx 9\%$ of the \upsi-triggered events have 2 or
more vertices.  For this analysis, we searched for candidates from
all the vertices found in an event.  We chose the vertex by
requiring that the vertex also matched the high momentum TPC tracks
that were already selected based on the BEMC tower extrapolation.
Since the BEMC is read out after every bunch crossing, out-of-bunch
pileup (interactions that happen before or after the triggered bunch
crossing) is rejected by the TPC track--BEMC cluster matching
requirement. We end up with a unique, unambiguous vertex in all
events. Within-bunch pileup (multiple interactions in the same bunch
crossing) is negligible for this analysis. Pileup rejection will
become more important as the luminosity of RHIC is increased.

Electrons were identified by selecting charged particle tracks with
specific ionization energy loss $dE/dx$ in the TPC consistent with
the ionization expected for electrons.  In the momentum region of
interest ($p \gtrsim 3$ \gevc), there is overlap between electrons
and charged pions, so a cut was placed that yielded an electron
efficiency of $\epsilon_{dE/dx}=84\%$. The cut was chosen to
optimize the effective signal $S_{\textrm{eff}}$ of single
electrons, $S_{\textrm{eff}}=S/(2B/S + 1)$, where $S$ is the
electron signal and $B$ is the hadron background.  To do this, we
construct a normal distribution of ionization measurements. We use
the measured ionization of each track, compare it to the expected
ionization for an electron and divide by the expected $dE/dx$
resolution (which depends on the track length and number of
measurements). This yields a normalized Gaussian distribution of
ionization measurements, $n\sigma_{dE/dx}$. We fit this distribution
with one Gaussian function to represent the electrons signal $S$,
and two Gaussian functions to model the background $B$ from pions
and other hadrons.

In addition, we used the combined information of the EMC energy $E$,
as obtained by the 3-tower candidate clusters from L2, and the TPC
track momentum $p$ to compute the $E/p$ ratio, which should be unity
for ultra-relativistic electrons to a very good approximation.
\begin{figure}[t]
\includegraphics*[width=\columnwidth]{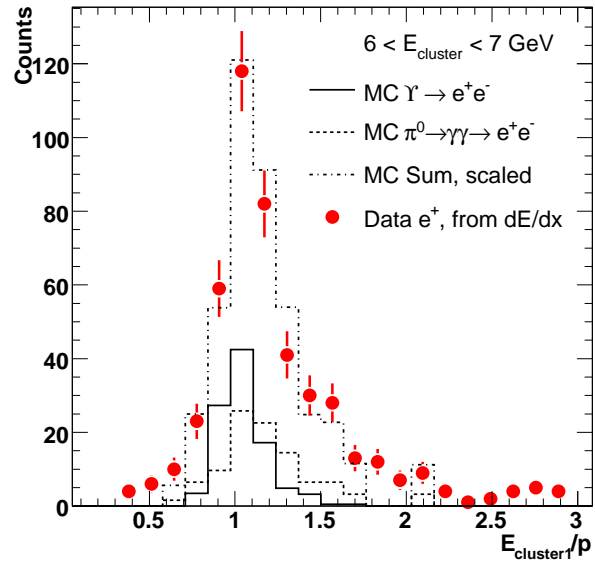}
\caption{\label{fig:EoverP} $E/p$ distribution from \upsi-triggered
data events, selecting a high-purity sample of positrons via $dE/dx$
(filled circles). We compare this to simulations from reconstructed
$e^{+}$ and $e^{-}$ from \upsi\ decays (solid line) and from $\pi^0$
events where the decay photons produce a conversion $e^{+}e^{-}$
pair (dashed line). The scaled sum of these two contributions is
shown as the dot-dashed line.}
\end{figure}
Figure~\ref{fig:EoverP} shows the $E/p$ distribution comparison from
an electron sample obtained from the Trigger I and II datasets
(filled circles).  Cuts were placed at $E/p=1.0 \pm 0.3$, which is
close to a 2$\sigma$ cut, given that fitting with a Gaussian gives
$\sigma=0.157 \pm 0.002$.  A shape that is approximately Gaussian is
expected for the $E/p$ distribution due to the Gaussian shape of the
resolution in both the energy measurement done in the BEMC and in
the curvature measurement done in the TPC (curvature is proportional
to $1/\pT$). We see that the $E/p$ distribution in the data shows a
non-Gaussian tail. We studied the shape in several ways. To rule out
distortions from misidentification of hadrons, we show only
positrons to avoid any antiproton contamination and use tight
$dE/dx$ cuts to select only a very high--purity electron sample. We
studied the energy dependence (to rule out threshold effects near 4
and 2.5 \gev), isolation cuts on the electron tracks (to rule out
contamination of the BEMC energy measurement from nearby charged
particles), and the $E/p$ shape of embedded electrons and photons.
We conclude that the tail at high $E/p$ has two sources:
\begin{enumerate}
\item Electron bremsstrahlung.
The solid line in Fig.~\ref{fig:EoverP} shows the simulated $E/p$
for \upsi\ electrons combined with a PYTHIA underlying \pp\ event,
embedded into zero-bias-triggered events.  Many of these produce a
Gaussian $E/p$ distribution, but we also see a tail. By selecting
simulated electrons which undergo bremsstrahlung, we find cases in
which the Monte Carlo momentum at the outermost TPC point differs
from the momentum at the \upsi\ decay vertex by more than 100 \mevc.
We then look at the $E/p$ of these electrons, and find that their
mean is shifted.  All entries in the region $E/p\approx1.3$ of the
solid line in Fig.~\ref{fig:EoverP} come from these cases.
Therefore, part of the non-Gaussian shape seen in the data is
accounted for by electron bremsstrahlung in the detector material,
and is included in the distribution of simulated \upsi\ daughters.

\item Photon conversions.
Events with a high-$Q^2$ di-jet that include back-to-back $\pi^0$'s
also fire the \upsi\ trigger.  Some of the $\pi^0$ daughter photons
will convert into \epluseminus\ pairs and leave a TPC track pointing
at the EMC cluster.  Generally, the $e^+$ and $e^-$ will strike the
calorimeter near the sibling photon, resulting in a track with a
high $E/p$ value.  We analyzed these simulated $\pi^0$ events and
applied the same tracking cuts, calorimeter clustering, and
BEMC--TPC matching used in this paper to the resulting electrons.
The $E/p$ distribution for the electrons in these simulated events
is shown as the dashed line in Fig.~\ref{fig:EoverP}, and we see a
non-Gaussian tail extending to larger values than that of the \upsi\
electrons. The average $E/p$ for the electrons from $\pi^0$ events
is $\approx1.18$ compared to $\approx1.08$ for \upsi\ electrons.
\end{enumerate}
Of these two effects, the first needs to be taken into account in
the \upsi\ efficiency because the \upsi\ daughters which undergo
bremsstrahlung and have a resulting $E/p$ value outside our cut will
be removed from the analysis. The second effect does not need to be
included in the $E/p$ efficiency since it is not due to bottomonium
events. These $\pi^0$--$\gamma$--conversion events will appear in
the invariant mass distributions, but they can be subtracted as
follows. A single photon conversion produces an unlike-sign pair.
When there are multiple photon conversions in an event, the
combinations that can be made include unlike-sign and like-sign
electron pairs. The unlike-sign pairs from a real photon will have
zero invariant mass and do not affect the analysis.  The additional
unlike-sign combinatorial pairs will have a distribution that can be
modeled by the like-sign combinations and are therefore removed when
subtracting the like-sign pair combinatorial background.

Since the $E/p$ distribution in the \upsi-triggered dataset has both
types of events, we reproduce the shape of the data using the two
types of electron simulations: the electron $E/p$ distributions from
\upsi\ events (including the bremsstrahlung tail) and from $\pi^0$
events. These are added and scaled to approximately fit the data in
the region $0.6 < E/p < 1.7$, shown as the dot-dashed line in
Fig.~\ref{fig:EoverP}.

As discussed above, we need to determine the efficiency for
electrons only from \upsi\ sources. To do this, we use the \upsi\
embedding simulations to estimate the efficiency of our $E/p$ cut,
which includes both the Gaussian shape expected from detector
resolution and the non-Gaussian tail due to bremsstrahlung. The
systematic uncertainty in the determination of the $E/p$ cut
efficiency is dominated by the uncertainty in the knowledge of the
detector material (and hence on the amount of bremsstrahlung), which
was estimated to be a factor of 2.  The material uncertainty is
different from that quoted in previous STAR electron analyses
because in this paper we do not restrict the event vertex to be near
the center of the detector, where the material budget is lower and
is better known. To estimate the uncertainty in the $E/p$ cut
efficiency, we construct an $E/p$ distribution by splitting the
original embedding simulation sample into electrons that undergo
bremsstrahlung and electrons that do not. We scale the number of
those that do by a factor of 2 and sum the two distributions to
obtain a new $E/p$ distribution.  The estimated efficiency found in
the bremsstrahlung-augmented distribution is lower by $\approx3\%$,
which we assign as the systematic uncertainty of $\epsilon_{E/p}$.
All the efficiencies used to obtain the \upsi\ cross section and
their systematic uncertainties are collected in
Table~\ref{tab:SystUncert}, below.

\section{Results and Discussion}\label{sec:ResultsDiscussion}
\begin{figure}
\includegraphics*[width=\columnwidth]{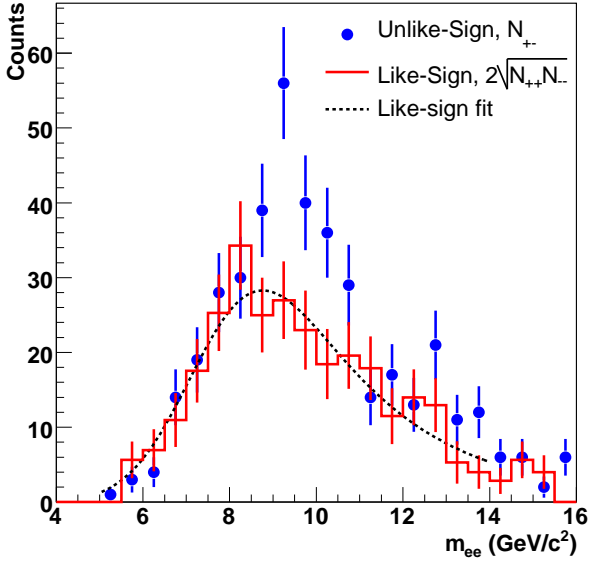}
\caption{\label{fig:upsilon_peakNoSub} Unlike-sign raw yield
$N_{+-}$ (filled circles) in the region $|y_{ee}|<0.5$, where
$y_{ee}$ is the pairwise rapidity, and like-sign combinatorial
background $2\sqrt{N_{++}N_{--}}$ (line histogram). The background
is fit to an exponential multiplied with an error function
representing the turn-on of the STAR \upsi\ trigger (dashed line).}

\end{figure}

Once we have tracks that satisfy our criteria and are matched to
tower clusters that satisfy all the L0 and L2 trigger requirements,
we form electron-positron pairs to produce the invariant mass
spectrum, using electron track momenta reconstructed in the TPC. We
use the like-sign combinations of $e^+e^+$ and $e^-e^-$ to estimate
the combinatorial background (see e.g.~\cite{Abreu:2000}) via
\begin{equation}\label{eq:LikeSignBackground}
    N^{\textrm{bck}}_{+-}=2\sqrt{N_{++}N_{--}}\cdot \frac{A_{+-}}{\sqrt{A_{++}A_{--}}}
\end{equation}
where $N_{+-}$ denotes the unlike-sign pair differential invariant
mass distribution $dN_{+-}/dm$ and $A_{+-}$ denotes the acceptance
for unlike-sign pairs (similarly for the like-sign distributions).
The symmetry of the BEMC and TPC for accepting unlike-sign and
like-sign pairs makes the ratio of acceptances unity, so we only use
$2\sqrt{N_{++}N_{--}}$ for the combinatorial background.  The
unlike-sign and like-sign background invariant mass spectra are
shown in Fig.~\ref{fig:upsilon_peakNoSub}.

\subsection{\upsi\ Line Shape}
In this analysis, we cannot resolve individual states of the
$\Upsilon$ family due to the limited statistics, finite momentum
resolution, and electron bremsstrahlung resulting from the large
material budget during the 2006 run. Therefore, the yield reported
here is for the combined \upsi(1S+2S+3S) states. The total
dielectron unlike-sign yield after like-sign background subtraction
has three contributions.
\begin{enumerate}
\item The \upsi\ states.
\item The Drell-Yan continuum.
\item The \bbbar\ continuum.
\end{enumerate}
In order to separate these three contributions, we obtain
parametrizations for the expected shapes of each of the three
components of the dielectron unlike-sign yield.  The contribution
from the \upsi\ states is obtained by the same simulations that were
used to obtain the detector efficiency.  The reconstructed invariant
mass shape for the 1S, 2S, and 3S states are individually obtained.
Each of the shapes from the simulation is fit with a functional form
introduced by the Crystal-Ball experiment~\cite{Gaiser:1982yw},
which can accommodate detector resolution and losses due to
bremsstrahlung in the detector material, and has the form:
\begin{equation}\label{eq:Crystal-Ball}
    f(m)=N \cdot \left\{  \begin{array}{cl}  \exp(-\frac{(m-\mu)^2}{2\sigma^2}), & \textrm{for } \frac{m-\mu}{\sigma}> -\alpha \\
                            A \cdot (B-\frac{m-\mu}{\sigma})^{-n},  & \textrm{for } \frac{m-\mu}{\sigma}\leq-\alpha
                          \end{array}
                \right.
\end{equation}
Requiring that the function and its derivative be continuous
constrains the constants $A$ and $B$ to be:
\begin{equation}\label{eq:Crystal-BallConstants}
\begin{array}{l}
A=\left(\frac{n}{|\alpha|}\right)^{n}\cdot\exp\left(-\frac{|\alpha|^2}{2}\right)\\
B=\frac{n}{|\alpha|}-|\alpha|
\end{array}
\end{equation}
We fix the parameters of the three Crystal-Ball functions
representing the three \upsi\ states and then adjust the relative
scales according to the average branching ratios and according to
the ratios of cross sections shown in
Table~\ref{tab:BranchRatiosAndCrossSec}. The overall integral of the
Crystal-Ball functions combined this way is left as a free parameter
in the fit, which allows us to obtain the \upsi\ yield.
\begin{figure}[t]
\includegraphics*[width=\columnwidth]{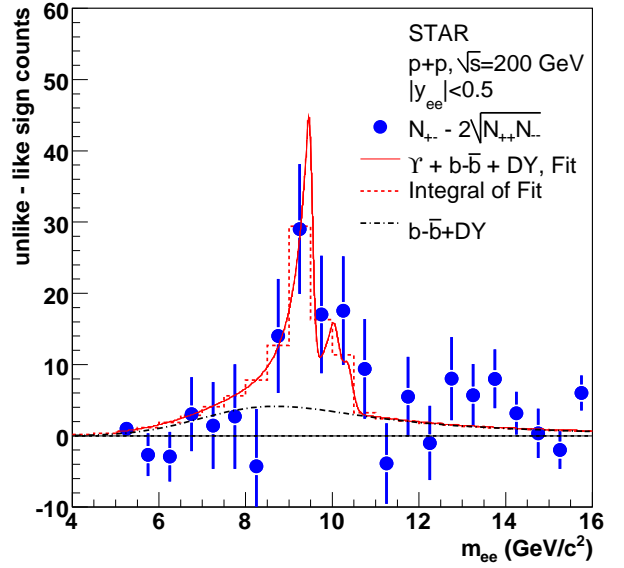}
\caption{\label{fig:upsilon_peak} The $e^+e^-$ signal after
subtracting the like-sign combinatorial background. The solid line
is the function used to fit the data, composed of: (a) three
Crystal-Ball functions (Eq.~\ref{eq:Crystal-Ball}) that represent
the combined $\upsi$(1S+2S+3S) line shape, (b) the Drell-Yan, and
(c) the \bbbar\ contributions. We fit the data using the integral of
the fit function in each mass bin, shown by the dashed-line
histogram. The sum of the two continuum contributions is shown by
the dot-dashed line. The integral of the Crystal-Ball functions
provides the net \upsi\ yield.}
\end{figure}

Figure \ref{fig:upsilon_peak} shows the data used to obtain the
\upsi\ and continuum yields.  The data points are the unlike-sign
\epluseminus\ signal after subtracting the like-sign combinatorial
background. The background subtraction is done bin-by-bin. The error
bars shown are statistical. The fit includes the contributions from
the \upsi(1S+2S+3S) states and the continuum contributions from
Drell-Yan and \bbbar\, shown as the solid-line function. The \upsi\
states are modeled with Eq.~\ref{eq:Crystal-Ball} to account for
detector resolution and for losses due to bremsstrahlung which
result in a tail to low values of invariant mass. The fit is done
using the integral of the fit function in each bin, and is shown as
the dashed-line histogram.  We discuss next the procedure used to
extract the continuum contributions (dot-dashed line) and the
\upsi(1S+2S+3S) yield.

\subsection{Drell-Yan and \bbbar\ Continuum
Contributions}\label{sec:DYandBBbar}

The Drell-Yan continuum is parameterized from a pQCD NLO
calculation~\cite{vogt} done in the kinematic range
$|y_{\Upsilon}|<0.5$ and $m>4$ \gevcc. We convolute the calculated
spectrum with the detector resolution (accounting for
bremsstrahlung), but this introduces very small changes in the shape
of the spectrum. We find that the shape of the Drell-Yan continuum
is well described by a function of the form
\begin{equation}\label{eq:Drell-YanParam}
\frac{A}{(1+m/m_{0})^{n}}
\end{equation}
with the parameters $m_{0}=2.70$ \gevcc\ and $n=4.59$.

The \bbbar\ contribution is parameterized from a simulation using
PYTHIA 8~\cite{Sjostrand:2008vc}.  We turn on production of \bbbar\
pairs, and follow their fragmentation and subsequent decays of the B
hadrons to look for dielectrons originating from the $b$ quarks.  We
convolute the simulated shape with the detector resolution and
bremsstrahlung, and find that the shape is well described by a
function of the form
\begin{equation}
\frac{A m^{b}}{(1+m/m_{0})^{c}}
\end{equation}
with the parameters $b=1.59$, $m_0=29.7$
\gevcc, and $c=26.6$.

Since the STAR \upsi\ trigger is meant to reject events with low
invariant mass, we parametrize the trigger response on both
continuum contributions by multiplying them with an error function
\begin{equation}
\frac{\textrm{erf}((m-m_{\mathrm{trig}})/w)+1}{2}
\end{equation}
where $m_{\mathrm{trig}}$ is related to the trigger thresholds and
$w$ describes the width of the turn-on of the error function due to
finite detector resolution. We obtain the parameters from the
like-sign data (Fig.~\ref{fig:upsilon_peakNoSub}) by multiplying the
error function with an exponential $\exp(-m/T)$ to account for the
random like-sign combinations at higher mass. We use the like-sign
data since the trigger turn-on shape for like-sign and unlike-sign
pairs is the same in our detector, but the like-sign invariant mass
distribution is purely due to combinatorics and can be fit with the
simple parametrization given above (Fig.
~\ref{fig:upsilon_peakNoSub}, dashed line). The parameters are
$m_{\mathrm{trig}}=8.1 \pm 0.8$ \gevcc\ and $w=1.8 \pm 0.5$ \gevcc.

To obtain the \upsi\ yields and cross section, we perform a fit of
the unlike-sign invariant mass distribution after subtracting the
like-sign background including contributions for the
\upsi(1S+2S+3S), and the continuum due to Drell-Yan and \bbbar.
Since the extracted \upsi\ yield from the fit will be sensitive to
the continuum yield, we next discuss the effect that variations of
the Drell-Yan and \bbbar\ cross sections can have on our result.

\begin{figure}[t]
\includegraphics*[width=\columnwidth]{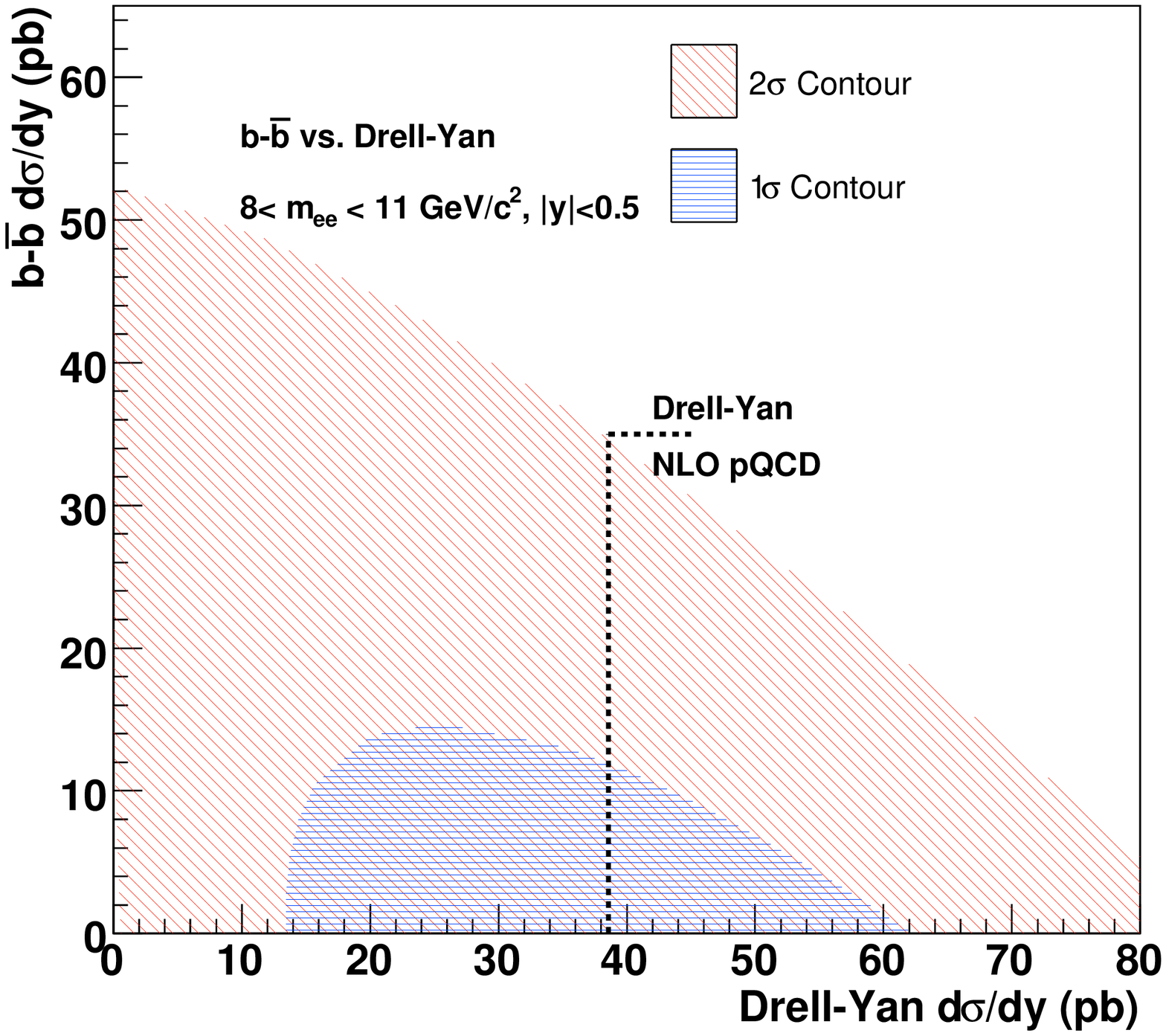}
\caption{\label{fig:chi2DYBBbar} The 1-$\sigma$ (horizontal lines)
and 2-$\sigma$ (diagonal lines) contours obtained by fitting the
data, showing the favored regions for the Drell-Yan and \bbbar\
cross sections. The dashed line shows the prediction for the
Drell-Yan cross section from a pQCD NLO calculation~\cite{vogt}.}
\end{figure}

We find that the resulting shapes of the two continuum contributions
are very similar.  We can fit the continuum yield using the
parameterized Drell-Yan and \bbbar\ contributions. The 1- and
2-$\sigma$ contours in the 2-D parameter space
$\sigma_{b\textrm{-}\bar{b}}$ \vs\ $\sigma_{\textrm{DY}}$ are shown
in Fig.~\ref{fig:chi2DYBBbar}, where the cross sections quoted are
corrected for efficiency and acceptance, and measured in the phase
space region $|y|<0.5$ and 8 $< m_{ee} <$ 11 \gevcc.

The fit shows a strong anti-correlation between the Drell-Yan and
the \bbbar\ cross sections due to the similarity of the
parameterized shapes. The shape of the Drell-Yan is slightly favored
by our data, and we obtain the minimal $\chi^2$ for a Drell-Yan
cross section of $\approx$ 38 pb with a negligible \bbbar\
contribution. The fit gives $\chi^2/\textrm{NDF}=1.1$. The
prediction from an NLO calculation of the Drell-Yan cross section
~\cite{vogt} is shown as the vertical line at 38.6 pb, and is
consistent with values in the 1-$\sigma$ region (horizontal lines).
Our data also allow values of the \bbbar\ cross section up to
$\approx$ 15 pb within the 1-$\sigma$ range, provided the Drell-Yan
yield is reduced. Given the anti-correlation, our data are mainly
sensitive to the sum of the two cross sections, and within the
1-$\sigma$ contour the allowed range for the sum of cross sections
is $(\sigma_{\textrm{DY}}+\sigma_{b\textrm{-}\bar{b}})|_{|y|<0.5,\
8<m<11\ \mathrm{GeV}/c^2} = 38 \pm 24$ pb.  This variation is taken
into account in the fit to extract the \upsi\ yield, and is included
in the quoted uncertainty.

\subsection{\upsi\ yield and cross section}\label{sec:UpsiYield}
The fitting function in Fig.~\ref{fig:upsilon_peak} includes a total
of three free parameters: the \upsi\ yield, the Drell-Yan yield and
the \bbbar\ yield.  This allows us to extract the \upsi\ yield
directly, automatically including ($i$) the statistical uncertainty
for each mass bin in Fig.~\ref{fig:upsilon_peak}, and ($ii$) the
uncertainty due to the anti-correlation between the continuum
contributions and the \upsi\ contribution.  To compare the yield
extraction using a different method, one can also perform bin
counting.  There are two ways to do this. One method is to take the
background-subtracted unlike-sign yield directly from the data shown
in Fig.~\ref{fig:upsilon_peak}, where the background subtraction is
done for each mass bin.  We then sum the resulting histogram in the
mass region of the \upsi\ peak, 8--11 \gevcc. This sum includes the
\upsi\ yield and the continuum contribution, so we subtract from it
the contribution from the Drell-Yan and \bbbar\ continuum obtained
in the fit.  A second method is to sum the yield of the unlike-sign
($N_{+-}$) in the region 8--11 \gevcc, do the same for the like-sign
positive ($N_{++}$), and like-sign negative ($N_{--}$), and then do
the subtraction $N_{+-}-2\sqrt{N_{++}N_{--}}$. In other words, in
the first method we do the subtraction bin-by-bin and then do the
sum, in the second method we do the sum first to get a single bin
and then we do the subtraction. The results for estimating
$N_{+-}-2\sqrt{N_{++}N_{--}}$ from the combined fit, the bin-by-bin
counting method, and the single-bin counting method, are shown in
Table~\ref{tab:UpsiYield}.
\begin{table}[tb]
\caption{\label{tab:UpsiYield} Extraction of \upsi(1S+2S+3S) yield
by bin counting and fitting. The sums are done in the range $8<m<11$
\gevcc.}

\begin{ruledtabular}
\begin{tabular}{lr}
    Fitting Results & \\
    $N_{+-}-2\sqrt{N_{++}N_{--}}$  & 80.9 \\
    \upsi\ counts  &   $59\pm 20$ \\
    \hline
    Bin-by-bin Counting & \\
    $N_{+-}-2\sqrt{N_{++}N_{--}}$  & 82.7 \\
    \upsi\ counts  &   $61\pm 20$ \\
    \hline
    Single-bin Counting & \\
    $N_{+-}-2\sqrt{N_{++}N_{--}}$ &  $75 \pm 20$ \\
    \upsi\ counts    & $54 \pm 18$ \\
\end{tabular}
\end{ruledtabular}
\end{table}

The quoted uncertainty in $N_{+-}-2\sqrt{N_{++}N_{--}}$ is listed
for the single-bin counting method, which is obtained from a
straightforward application of the statistical errors of the
corresponding yields ($N_{+-}=230$, $N_{++}=92$, $N_{--}=65$). It
amounts to a contribution of 26\% to the uncertainty, and originates
only from counting statistics. In addition to this purely
statistical uncertainty, we must also take into account the
uncertainty of the continuum subtraction.  This part of the
uncertainty is obtained from the fitting method in all the quoted
\upsi\ counts in the Table.  It should be noted that one cannot use
counting statistics to estimate the uncertainty on the \upsi\ yield
due to the continuum subtraction. The reason is that the \upsi\ and
continuum yields are obtained from the same data, namely the
invariant mass dielectron spectrum in Fig.~\ref{fig:upsilon_peak}.
The two yields are therefore anti-correlated: a larger continuum
yield reduces the extracted \upsi\ counts.  One must take this
anti-correlation into account in the estimation of the uncertainty
on the \upsi\ yield. The best way to do this is to use the fitting
method.  In the fit, both the statistical precision of the invariant
mass spectrum and the anti-correlation between the \upsi\ and
continuum yields in the estimation of the uncertainty are taken into
account when varying the \upsi\ yield and the continuum yield to
minimize the $\chi^2$.   We find the uncertainty from the fit to be
33\% of the \upsi\ yield, and we hence use this fitting-method
result to estimate the uncertainty in all the \upsi\ yields quoted
in Table~\ref{tab:UpsiYield}.  We quote this as the ``stat. + fit''
uncertainty in our cross section result.  It should be noted that
with the statistics of the present analysis, we find that the
allowed range of variation of the continuum yield in the fit is
still dominated by the statistical error bars of the invariant mass
distribution, and so the size of the 33\% uncertainty is mainly
statistical in nature.  However, we prefer to denote the uncertainty
as ``stat. + fit'' to clarify that it includes the estimate of the
anti-correlation between the \upsi\ and continuum yields obtained by
the fitting method.   A systematic uncertainty due to the continuum
subtraction can be estimated by varying the model used to produce
the continuum contribution from \bbbar.  These variations produce a
negligible change in the extracted yield with the current
statistics.

When comparing different counting methods in
Table~\ref{tab:UpsiYield}, we see that the difference between the
bin-by-bin counting method and the fitting method is negligible. The
single-bin counting method yield is lower than the one from the
bin-by-bin counting method by 9\%, and we assign this as a component
of the systematic uncertainty.

Table~\ref{tab:UpsiYield} also lists the yield of \upsi(1S+2S+3S).
In the fitting method, the yield and the quoted uncertainty is
obtained directly from the fit, as well as the contribution from
Drell-Yan and \bbbar\ discussed in the previous section.  This has
the advantage that the uncertainty due to the continuum subtraction
and the correlations between the \upsi, Drell-Yan and \bbbar\
contributions are automatically taken into account when exploring
the parameter space.  The disadvantage is that there is a model
dependence on the line-shapes used for the fit.  In the counting
methods, to extract the \upsi\ yield we subtract the continuum
contribution obtained from the fit.  This reduces the model
dependence on the \upsi\ line shape.  However, we must still account
for the uncertainty in the estimate of the continuum contribution in
the determination of the \upsi\ yield uncertainty.  Since this
uncertainty should include similar correlations between the \upsi\
yield and continuum yields as found in the fitting method, the
relative uncertainties should be approximately equal.  Therefore, we
use the same relative \upsi\ yield uncertainty for the counting
methods as for the fitting method.  To get the total \upsi\ yield we
must correct the above numbers for the yield outside the integration
region. This correction can be obtained from the fitted Crystal-Ball
functions, and gives an additional 9\% contribution to the \upsi\
yield. We report results for the cross section using the bin-by-bin
counting method.

In order to transform the measured yield of \upsi(1S+2S+3S) into a
cross section, we applied several correction factors:
\begin{eqnarray}
\sum_{n=1}^3 \mathcal{B}(n\textrm{S})\times \sigma(n\textrm{S})=
\frac{N}{\Delta y\times\epsilon\times\mathcal{L}},
\label{eq:upsilon}
\end{eqnarray}
where the symbols are as follows. $\mathcal{B}(n\textrm{S})$ is the
branching fraction for $\Upsilon(n\textrm{S})\rightarrow e^+e^-$.
$\sigma(n\textrm{S})$ is the cross section $d\sigma/dy$ for the $n$S
state in the region $|y_{\Upsilon}|<0.5$. $N=67\pm 22$ (stat.) is
the measured \upsi(1S+2S+3S) yield from the bin-by-bin counting
method in Table~\ref{tab:UpsiYield} with a 9\% correction to account
for the yield outside $8< m_{ee} < 11$ \gevcc.  $\Delta y=1.0$ is
the rapidity interval for our kinematic region $|y_{\Upsilon}|<0.5$.
The total efficiency for reconstructing members of the \upsi\ family
is the product
$\epsilon=\epsilon_{\textrm{geo}}\times\epsilon_{\textrm{vertex}}
\times\epsilon_{\textrm{L0}} \times\epsilon_{\textrm L2}
\times\epsilon_{\textrm{TPC}} \times\epsilon_{R}
\times\epsilon_{dE/dx} \times\epsilon_{E/p}$, where the symbols are
as follows. $\epsilon_{\textrm{geo}}$ is the BEMC geometrical
acceptance. $\epsilon_{\textrm{vertex}}$ is the vertex-finding
efficiency. $\epsilon_{\textrm{L0}}$ and $\epsilon_{\textrm{L2}}$
are the trigger efficiencies for L0 and L2, respectively.
$\epsilon_{\textrm{TPC}}$ is the TPC geometrical acceptance times
tracking efficiency for reconstructing both daughters in the TPC.
$\epsilon_{R}$ is the TPC-BEMC $\eta$-$\varphi$ matching efficiency.
$\epsilon_{dE/dx}$ is the electron identification efficiency from
the specific ionization requirement, and $\epsilon_{E/p}$ is the
electron identification efficiency from the $E/p$ selection.

We find for the cross section at midrapidity in $\sqrt{s}=200$ GeV
\pp\ collisions the result
\begin{eqnarray}
\label{eq:UpsilonSigmaResult} \sum_{n=1}^3
\mathcal{B}(n\textrm{S})\times \sigma(n\textrm{S})= 114\pm 38\
^{+23}_{-24}\ \textrm{pb}\ .
\end{eqnarray}
The uncertainties quoted are the 33\% statistical+fit uncertainty
(mentioned in the discussion of Table~\ref{tab:UpsiYield}) and the
systematic uncertainty, respectively.

The major contributions to the systematic uncertainty are: the
uncertainty in the choice of bin-counting method, the uncertainty in
the integrated luminosity, the uncertainty in the BBC efficiency for
\pp\ NSD events and the uncertainty in the L0 trigger efficiency for
\upsi\ events. The polarization of the \upsi\ states also affects
the estimation of the geometrical acceptance. We estimate this
uncertainty by comparing simulations of fully longitudinal and fully
transverse decays and comparing the acceptance of these cases with
the unpolarized case. A list of all corrections and systematic
uncertainties in the procedure to extract the cross section is
compiled in Table \ref{tab:SystUncert}. The combined systematic
uncertainty is obtained by adding all the sources in quadrature.
Note that the single-particle efficiencies enter quadratically when
reconstructing dielectron pairs, so we multiply the single-particle
uncertainty by a factor of 2 when estimating the pairwise
uncertainty.

\begin{table}[tb]
\caption{\label{tab:SystUncert} Systematic uncertainties on the
measurement of the \upsi\ cross section.}

\begin{ruledtabular}
\begin{tabular}{ccc}
    Quantity & Value & Syst. uncertainty on $d\sigma/dy$ (\%)\\
    \hline
  $N_{+-}-2\sqrt{N_{++}N_{--}}$       & 82.7  & $^{+0}_{-9}$ \\
  $\mathcal{L}$        & 7.9 pb$^{-1}$  & $\pm 7$ \\
  $\epsilon_{\mathrm{BBC}}$   & 0.87         & $\pm 9$ \\
  $\epsilon_{\mathrm{geo}}$   & 0.57          & $^{+3.0}_{-1.7}$ \\
  $\epsilon_{\textrm{vertex}}$ & 0.96         & $\pm 1.0$ \\
  $\epsilon_{\textrm{L0}}$    & 0.43          & $^{+7.5}_{-5.9}$ \\
  $\epsilon_{\textrm{L2}}$    & 0.85          & $^{+0.7}_{-0.2}$ \\
  $\epsilon_{\textrm{TPC}}$   & 0.85$^{2}$      & $2\times \pm 5.8$ \\
  $\epsilon_{R}$              & 0.93$^{2}$      & $2\times ^{+1.1}_{-0.2}$ \\
  $\epsilon_{dE/dx}$          & 0.84$^{2}$     & $2\times \pm 2.4$ \\
  $\epsilon_{E/p}$            & 0.93$^{2}$     & $2\times \pm 3.0$\\
  \hline
  \\
  Combined                    &               & $^{+22.8}_{-24.1}$  pb\\
\end{tabular}
\end{ruledtabular}
\end{table}
\begin{figure}[htb]
\includegraphics*[width=\columnwidth]{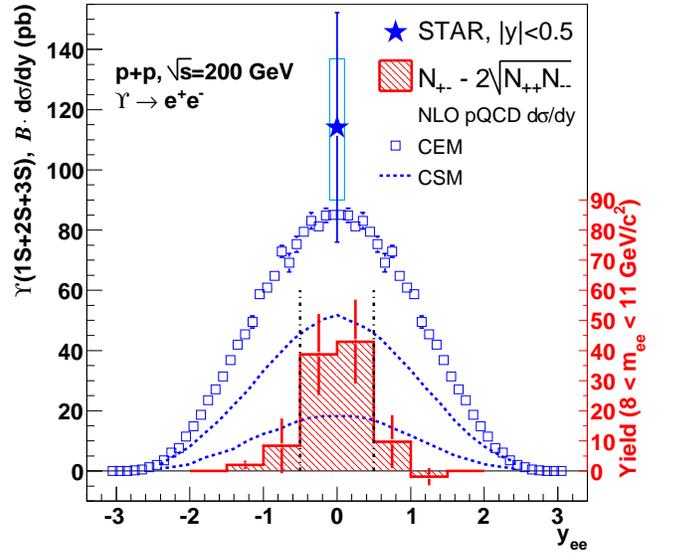}
\caption{\label{fig:upsilon_cross_section} The STAR measurement of
the midrapidity \upsi(1S+2S+3S) cross section times branching ratio
into electrons (star). Error bars are statistical, the box shows the
systematic uncertainty, and the scale is given by the left axis. The
raw yield vs. $y$ is shown by the histogram at the bottom
(diagonal-line fill pattern), with scale on the right axis.  The
cross section was calculated from the yield between the vertical
dot-dashed lines, $|y_{ee}|<0.5$. The open squares are from an NLO
CEM calculation, and the two dotted lines give the limits for the
prediction from a NLO CSM calculation of the \upsi\ cross section
(see text).}
\end{figure}

The result we obtain for the cross section is shown in
Fig.~\ref{fig:upsilon_cross_section}, where the datum point given by
the star symbol is our measurement, the error bars and the box
depict the statistical and systematic uncertainties, respectively.
To illustrate the acceptance in rapidity, we also show the
unlike-sign pairs after like-sign background subtraction,
$N_{+-}-2\sqrt{N_{++}N_{--}}$, in the \upsi\ region $8<m_{ee}<11$
\gevcc\ as a hashed histogram.  The scale on the right axis of the
figure is used for the counts in the histogram, and the scale in the
left axis of the figure is used for the cross section. We compare
our measurement with NLO CEM predictions~\cite{vogt} of the
\upsione\ rapidity distribution. Since we measure all three states
and only in the dielectron channel, the calculation of the \upsione\
is scaled by a factor
\begin{equation}
\frac{\mathcal{B}\textrm{(1S)}\times\sigma\textrm{(1S)}+\mathcal{B}\textrm{(2S)}\times\sigma\textrm{(2S)}
    +\mathcal{B}\textrm{(3S)}\times\sigma\textrm{(3S)}}
    {\sigma\textrm{(1S)}}
\end{equation}
in order to compare it to our measurement of the cross section for
all three states. The branching ratios and cross sections used for
this scale factors are those from
Table~\ref{tab:BranchRatiosAndCrossSec}.
The calculation is in agreement with our measurement. The two dotted
lines in the plot are the upper and lower bounds of the cross
section obtained from a calculation in the CSM for direct \upsione\
production~\cite{Brodsky:2009cf} based on NLO code developed for
quarkonium production at hadron colliders~\cite{Campbell:2007ws}.
Since the calculation is for the 1S state alone and for direct
\upsi\ production (ignoring feed-down from P-states), to compare to
our measurement, which includes all 3 states and feed-down
contributions, the values from the calculation were divided by a
factor 0.42 to account for this (see Ref~\cite{Brodsky:2009cf} for
details). The bounds in the calculation are obtained by varying the
bottom quark mass and the renormalization and factorization scales.
The CSM prediction is lower than our data, indicating that
additional contributions are needed beside production via color
singlet.
\begin{figure}[tb]
\includegraphics*[width=\columnwidth]{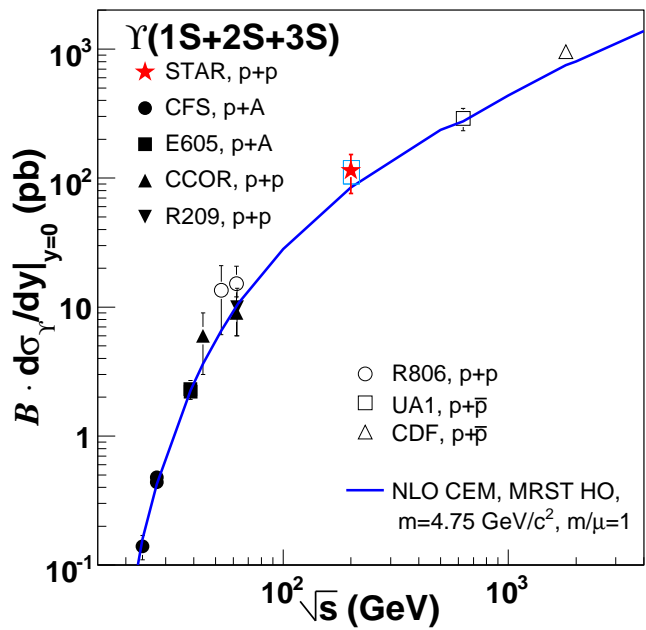}
\caption{\label{fig:upsilon_world_data} Evolution of the
\upsi(1S+2S+3S) cross section with center-of-mass energy for the
world data and an NLO CEM calculation. The error bars on the STAR
datum point are statistical and systematic as in
Fig.~\ref{fig:upsilon_cross_section}.}
\end{figure}

In Fig.~\ref{fig:upsilon_world_data}, we also compare our
\upsi(1S+2S+3S) result with measurements done in $p+A$, \pp\ and
$p+\bar{p}$ collisions at center-of-mass energies ranging from 20
GeV up to 1.8 TeV
\cite{Yoh:1978id,Ueno:1978vr,Childress:1985fd,Yoshida:1989pi,Moreno:1990sf,Camilleri:1979qf,
Kourkoumelis:1980hg,Angelis:1979ar,Albajar:1986iu,Acosta:2001gv},
and to NLO CEM predictions~\cite{rhic2} for a wide range of
center-of-mass energies.

Our result is consistent with the overall trend, and provides a
reference for bottomonium production at the top RHIC energy.

\section{Conclusions}\label{sec:Conclusions}
The STAR experiment has measured the
$\Upsilon\textrm{(1S+2S+3S)}\rightarrow e^+e^-$ cross section at
midrapidity, $|y_{ee}|<0.5$, in \pp\ collisions at $\sqrt{s}=200$
GeV to be $(\mathcal{B}\times d\sigma/dy)^{\textrm{1S+2S+3S}}=114\pm
38~\textrm{(stat. + fit)} ^{+23}_{-24}~\textrm{(syst.)}$ pb.
Calculations done in the Color Evaporation Model at NLO are in
agreement with our measurement, while calculations in the Color
Singlet Model underestimate our cross section by $\approx2\sigma$.
Our result is consistent with the trend as a function of
center-of-mass energy based on data from other experiments. We
report a combined continuum cross section, Drell-Yan plus \bbbar $\
\rightarrow \epluseminus$, measured in the kinematic range
$|y_{ee}|<0.5$ and $8<m_{ee}<11$ \gevcc, of
$(\sigma_{\textrm{DY}}+\sigma_{b\textrm{-}\bar{b}}) = 38 \pm 24$ pb.
The STAR measurement presented here will be used as a baseline for
studying cold and hot nuclear matter effects in \dAu\ and \AuAu\
collisions, as the relatively clean environment provided by the STAR
high-mass dielectron trigger permits the approach outlined in this
paper to be deployed up to the most central \AuAu\ collisions. With
increased luminosity, a better determination of the cross section,
its \pT\ dependence and a separation of the 2S and 3S states will be
possible. The projected luminosity upgrades to RHIC should increase
the $\Upsilon$ yield to $\approx8 300$ in \pp\ and $\approx11 200$
in \AuAu\ collisions during one RHIC year~\cite{rhic2}. The
increased statistics will greatly reduce the uncertainty in the
determination of the continuum cross section and will allow a
thorough study of the bottomonium sector by resolving the 2S and 3S
states.

The authors thank R. Vogt and J.-P. Lansberg for providing several
calculations and for useful discussions. We thank the RHIC
Operations Group and RCF at BNL, the NERSC Center at LBNL and the
Open Science Grid consortium for providing resources and support.
This work was supported in part by the Offices of NP and HEP within
the U.S. DOE Office of Science, the U.S. NSF, the Sloan Foundation,
the DFG cluster of excellence `Origin and Structure of the Universe'
of Germany, CNRS/IN2P3, STFC and EPSRC of the United Kingdom, FAPESP
CNPq of Brazil, Ministry of Ed. and Sci. of the Russian Federation,
NNSFC, CAS, MoST, and MoE of China, GA and MSMT of the Czech
Republic, FOM and NWO of the Netherlands, DAE, DST, and CSIR of
India, Polish Ministry of Sci. and Higher Ed., Korea Research
Foundation, Ministry of Sci., Ed. and Sports of the Rep. Of Croatia,
Russian Ministry of Sci. and Tech, and RosAtom of Russia.


\end{document}